\documentclass[a4paper]{jpconf}
\usepackage{graphicx}

 \usepackage{amssymb,amsfonts}
\usepackage{bm}
\usepackage{amsmath}
 \usepackage{color}
 \usepackage{appendix}
 \usepackage{amsthm}
 \usepackage{titlesec}

\renewcommand{\vec}[1]{\mathbf{#1}}
\newcommand{\functional}[1]{\mathcal{#1}}

\newcommand{\transpose}{\mathrm{t}}

\newcommand{\complexi}{\mathrm{i}}

\newcommand{\ignore}[1]{}

\begin{document}

\title{Multimode Bogoliubov transformation and Husimi's Q-function}

\author{Joonsuk Huh$^{1,2}$}

\address{$^1$Department of Chemistry, Sungkyunkwan University, Suwon 16419, Korea}
\address{$^2$SKKU Advanced Institute of Nanotechnology (SAINT), Sungkyunkwan University, Suwon 16419, Korea}
\ead{joonsukhuh@gmail.com}


\begin{abstract}
In this paper, we present numerical schemes for evaluating the matrix elements of Gaussian/non-Gaussian operators in the Fock state basis, which are identified as multivariate Hermite polynomials (MHPs). Using the integral transformation operator to perform the multimode Bogoliubov transformation, Husimi's Q-functions of Gaussian/non-Gaussian operators are easily derived as the generating functions of MHPs. 
\end{abstract}

%
%
%
%
%

\section{Introduction}
 Multivariate Hermite polynomials (MHPs)
 play important roles in various field of research, including quantum optics~\cite{dondonov1994,kok2001} and   molecular spectroscopy~\cite{doktorov:1977,jankowiak:2007}. However, the evaluation of MHPs is notoriously difficult, and various algorithms have been developed for this task, such as Willink's recursion formula~\cite{willink:2005}. Willink~\cite{willink:2005} also showed that the MHPs could be converted to multivariate Gaussian moments (MGMs). Kan~\cite{kan:2007} developed an efficient direct summation formula for MGMs. In his paper, the connection between the MGMs and the hafnian~\cite{hamilton2017,kruse2019,kamil2018,quesada2019} is also given. Therefore, the MHPs, MGMs, and the hafnian are mutually convertible quantities that have the same computational complexity, but are labeled differently.        
 
 In molecular vibronic spectroscopy, the vibronic transition amplitudes between the two different electronic potential energy surfaces (represented as harmonic potential wells) are called the Franck-Condon integral (FCI), which is usually evaluated as MHPs. Although all MHP parameters are real values, the computational difficulties have led to various numerical strategies~\cite{jankowiak:2007,Huh2011a} developed in computational chemistry packages. Doktorov et al.~\cite{Doktorov1975} interpreted the molecular vibronic transition as a Gaussian quantum optical transformation-- the multimode Bogoliubov transformation. The authors decomposed the vibronic transition operation into elementary quantum optical operators (displacement, rotation, and squeezing). Proceeding from the work of Doktorov et al.~\cite{Doktorov1975}, after 40 years, Huh et al.~\cite{huh2015,Huh2017VBS} connected the story of molecular vibronic spectroscopy to Gaussian boson sampling~\cite{rahimi2015,hamilton2017,kruse2019}, which is a quantum sampling problem of Gaussian light. Some experiments were performed successfully by quantum devices at the small scale~\cite{Shen2018,Clements_2018,wang2019}. The properties of Gaussian states are well explained in Refs.~\cite{Weedbrook2012,Adesso2014}, and the detailed calculations involving Gaussian/non-Gaussian operations can be found in Refs.~\cite{gagatsos2019,quesada2019b,su2019}. Therein, Husimi's Q-function~\cite{husimi} for the Gaussian light was derived, as a generating function, from Wigner's function by convolution. Only a few papers ~\cite{dondonov1994} have shown the complex Bogoliubov parameter dependence on Husimi's Q-function explicitly.      

In this paper, we connect the matrix elements of non-Gaussian operators in the Fock state basis to the MHPs so that the evaluation can be conducted in an equal footing as the Gaussian case. Unlike the recent papers~\cite{gagatsos2019,quesada2019b,su2019}, we start from the multimode Bogoliubov transformation to define the pure Gaussian operator. We adapt the integral operator method of Fan et al.~\cite{Hong-yi1994,Fan2003} and follow the derivation steps in Ref.~\cite{Hong-yi1994} throughout the paper to obtain Husimi's Q-function, which is used as a generating function of the MHPs for the Gaussian/non-Gaussian matrix elements in the Fock state basis. With the aid of the integral operator method proposed by Fan et al., the explicit parameter dependence on the multimode Bogoliubov transformation matrices is made in Husimi's Q-function. 

Our work is presented as follows: The relationship between the matrix elements of Gaussian/non-Gaussian operators in the Fock state basis, the MHPs, and the multivariate Gaussian distribution is given first (some contents are taken from the PhD thesis (Ch. 3) of the author~\cite{Huh2011a} for the following section). Then, the integral operator transformation~\cite{Hong-yi1994,Fan2003} method is introduced and applied to the multimode Bogoliubov transformation for the corresponding Husimi's Q-function of Gaussian/non-Gaussian operators. Finally, we summarize the paper in the conclusion section.  

\section{Matrix elements of Gaussian/non-Gaussian operator, multivariate Hermite polynomials, and moments of multivariate Gaussian distribution}
The matrix elements of a multimode Gaussian operator ($\hat{O}_{G}$) in the $N$-dimensional Fock states ($\vert\mathbf{n}\rangle=\vert n_{1},\cdots,n_{N}\rangle$ and $\vert\mathbf{m}\rangle=\vert m_{1},\cdots,m_{N}\rangle$) can be generated by Husimi's Q-function of a Gaussian operator ($\pi^{-N}\langle\boldsymbol{\alpha}\vert\hat{O}_{G}\vert\boldsymbol{\alpha}\rangle$)~\cite{husimi}; this can easily be seen via the expansion of the coherent states ($\vert\boldsymbol{\alpha}\rangle$) in the Fock state basis, as follows, with the unnormalized Husimi Q-function,  
\begin{align}
 H_{G}(\hat{O}_{G},\boldsymbol{\alpha})&=\exp(\boldsymbol{\alpha}^{\dagger}\boldsymbol{\alpha})\langle \boldsymbol{\alpha} \vert\hat{O}_{\mathrm{G}}\vert \boldsymbol{\alpha}\rangle,\nonumber\\ 
 &=\sum_{\mathbf{n},\mathbf{m}=\mathbf{0}}^{\boldsymbol{\infty}}\prod_{j=1}^{N}\left(\frac{\alpha_{j}^{n_{j}}(\alpha_{j}^{*})^{m_{j}}}{\sqrt{n_{j}!m_{j}!}}\right)\langle\mathbf{m}\vert\hat{O}_{G}\vert\mathbf{n}\rangle. 
\end{align}
Therefore, the matrix elements in the Fock basis are given as partial derivatives with respect to the phase parameters of the coherent states, \emph{i.e.}  
\begin{equation}
\langle \mathbf{m} \vert\hat{O}_{\mathrm{G}}\vert \mathbf{n}\rangle=\prod_{j=1}^{N}\frac{\partial_{\alpha_{j}}^{n_{j}}\partial_{\alpha_{j}^{*}}^{m_{j}}}{\sqrt{n_{j}!m_{j}!}}H_{G}(\hat{O}_{G},\boldsymbol{\alpha})\Big\vert_{\boldsymbol{\alpha}=\mathbf{0}}.
\end{equation}
Here, the small and large letters in bold fonts are used to indicate column vectors and square matrices, respectively.

The unnormalized Husimi Q-function of the Gaussian operator ($\exp(\boldsymbol{\alpha}^{\dagger}\boldsymbol{\alpha})\langle\boldsymbol{\alpha}\vert\hat{O}_{G}\vert\boldsymbol{\alpha}\rangle$) has a closed form of the multivariate Gaussian distribution function, 
\begin{align}
H_{\mathrm{G}}(\mathbf{V},\boldsymbol{\mu})=\langle\mathbf{0}\vert\hat{O}_{G}\vert\mathbf{0}\rangle
\exp\Big(-\frac{1}{2}\bar{\boldsymbol{\alpha}}^{\mathrm{T}}
\mathbf{V}    
\bar{\boldsymbol{\alpha}}+
    \boldsymbol{\mu}^{\mathrm{T}}
    \bar{\boldsymbol{\alpha}}
    \Big),     
\end{align}
where $\bar{\boldsymbol{\alpha}}^{\mathrm{T}}=(\boldsymbol{\alpha}^{\mathrm{T}} \boldsymbol{\alpha}^{\dagger})$.  
The matrix elements are given as the MHPs $\mathcal{H}$, 
\begin{align}
    \langle \mathbf{m} \vert\hat{O}_{\mathrm{G}}\vert \mathbf{n}\rangle&=\langle\mathbf{0}\vert\hat{O}_{G}\vert\mathbf{0}\rangle
    \prod_{j=1}^{N}\left(\frac{(-1)^{(n_{j}+m_{j})}}{\sqrt{n_{j}!m_{j}!}}\right)\nonumber\\ 
    &~~~\exp\left(-\tfrac{1}{2}(\bar{\boldsymbol{\alpha}}-\mathbf{V}^{-1}\boldsymbol{\mu})^{\mathrm{T}}\mathbf{V}(\bar{\boldsymbol{\alpha}}-\mathbf{V}^{-1}\boldsymbol{\mu})\right)
    \mathcal{H}_{\bar{\boldsymbol{\alpha}}}(\bar{\boldsymbol{\alpha}}-\mathbf{V}^{-1}\boldsymbol{\mu};\mathbf{V}^{-1})\Big\vert_{\bar{\boldsymbol{\alpha}}=\mathbf{0}}, \nonumber\\ 
    &=\langle\mathbf{0}\vert\hat{O}_{G}\vert\mathbf{0}\rangle
    \prod_{j=1}^{N}\left(\frac{(-1)^{(n_{j}+m_{j})}}{\sqrt{n_{j}!m_{j}!}}\right)\mathcal{H}_{\bar{\boldsymbol{\alpha}}}(-\mathbf{V}^{-1}\boldsymbol{\mu};\mathbf{V}^{-1}),\nonumber\\  
    &=\langle\mathbf{0}\vert\hat{O}_{G}\vert\mathbf{0}\rangle
    \prod_{j=1}^{N}\left(\frac{1}{\sqrt{n_{j}!m_{j}!}}\right)\mathcal{H}_{\bar{\boldsymbol{\alpha}}}(\mathbf{V}^{-1}\boldsymbol{\mu};\mathbf{V}^{-1}),
    \label{eq:MEF}
\end{align}
where the $M$-dimensional Hermite polynomials are defined as follows, 
\begin{equation}
\mathcal{H}_{\mathbf{v}}(\mathbf{x};\boldsymbol{\Lambda})=
(-1)^{\tilde{v}}\exp(\tfrac{1}{2}\mathbf{x}^{\mathrm{T}}\boldsymbol{\Lambda}^{-1}\mathbf{x})
\prod_{j=1}^{M}(\partial_{x_{j}}^{v_{j}})\exp(-\tfrac{1}{2}\mathbf{x}^{\mathrm{T}}\boldsymbol{\Lambda}^{-1}\mathbf{x}), 
\end{equation}
where $\tilde{v}=\sum_{j=1}^{M}v_{j}$, with a complex symmetric matrix $\boldsymbol{\Lambda}$ having a symmetric positive definite real part. 
We can evaluate the MHPs using the following recursion relation, 
\begin{align}
 \mathcal{H}_{v_{1},\cdots,v_{k}+1,\cdots,v_{M}}(\mathbf{x};\boldsymbol{\Lambda})=
 \left[\sum_{j=1}^{M}(\boldsymbol{\Lambda}^{-1})_{kj}x_{j}\right]\mathcal{H}_{\mathbf{v}}(\mathbf{x};\boldsymbol{\Lambda})-\sum_{j=1}^{M}(\boldsymbol{\Lambda}^{-1})_{kj}v_{j}\mathcal{H}_{v_{1},\cdots,v_{j}-1,\cdots,v_{M}}(\mathbf{x};\boldsymbol{\Lambda}),
 \label{eq:hermiterec}
\end{align}
which was derived by Willink~\cite{willink:2005}. 
Willink also found the recursion relation for the MHPs via the relation between the multivariate normal (Gaussian) moments $\mathcal{E}$ and the MHPs $\mathcal{H}$, \emph{i.e.} 
\begin{equation}
    \mathcal{E}(\prod_{j=1}^{M}y_{j}^{v_{j}})=\mathrm{i}^{-\tilde{v}}\mathcal{H}_{\mathbf{v}}(\mathrm{i}\boldsymbol{\Lambda}\mathbf{y_{m}};\boldsymbol{\Lambda}),
\end{equation}
where $\vec{y}$ is a random variable vector in a multivariate normal distribution,  
$\functional{N}(\vec{y_{m}},\boldsymbol{\Lambda}^{-1})$  
with its mean vector $\vec{y_{m}}$ and covariance matrix
$\boldsymbol{\Lambda}^{-1}$.
Exploiting the 
Magnus series expansion for products of
variables: 
\begin{equation}
\prod_{k}^{M}y_{k}^{v_{k}}
=(\tilde{v}!)^{-1}\sum_{l_{1}=0}^{v_{1}}\cdots\sum_{l_{M}=0}^{v_{M}}
(-1)^{\sum_{k}^{M}l_{k}}
\begin{pmatrix}v_{1} \\ l_{1}
\end{pmatrix}
\cdots
\begin{pmatrix}v_{M} \\ l_{M}
\end{pmatrix}
(\vec{h}^\transpose\vec{y})^{\tilde{v}},
\label{eq:etoh}
\end{equation}
where
$\vec{h}=\vec{v}/2-\vec{l}$, 
Kan~\cite{kan:2007} developed an efficient algorithm to evaluate the
MGM, 
\begin{align}
\functional{E}(\prod_{k}^{M}y_{k}^{v_{k}})
=\sum_{l_{1}=0}^{v_{1}}\cdots\sum_{l_{M}=0}^{v_{M}}
\sum_{s=0}^{[\tilde{v}/2]}(-1)^{\sum_{k}^{M}l_{k}}
\begin{pmatrix}v_{1} \\ l_{1}
\end{pmatrix}
\cdots
\begin{pmatrix}v_{M} \\ l_{M}
\end{pmatrix}
\frac{\Big(\tfrac{\vec{h}^{\transpose}\boldsymbol{\Lambda}^{-1}\vec{h}}{2}
\Big)^{s}
(\vec{h}^{\transpose}\vec{y_{m}})^{\tilde{v}-2s}}{s!(\tilde{v}-2s)!},
\label{eq:imoments}
\end{align}
here 
 $[\tilde{v}/2]$ is the greatest integer less than or equal to $\tilde{v}/2$. 
An (alternative) iterative evaluation scheme for the
MHPs~\cite{Huh2011a},  
may be evaluated efficiently by identifying
$\vec{y_{m}}=-\complexi\boldsymbol{\mu}$ and 
$\boldsymbol{\Lambda}=\mathbf{W}^{-1}$  
in Eqs.~\eqref{eq:etoh} and~\eqref{eq:imoments}, \emph{i.e.} 

\begin{align}
\functional{H}_{\vec{v}}(\mathbf{W}^{-1}\boldsymbol{\mu};\mathbf{W}^{-1})=
\sum_{l_{1}=0}^{v_{1}}\cdots\sum_{l_{M}=0}^{v_{M}}
\sum_{s=0}^{[\tilde{v}/2]}(-1)^{\sum_{k}^{M}l_{k}+s}
\begin{pmatrix}v_{1} \\ l_{1}
\end{pmatrix}
\cdots
\begin{pmatrix}v_{M} \\ l_{M}
\end{pmatrix}
\frac{\Big(\tfrac{\vec{h}^{\transpose}\mathbf{W}\vec{h}}{2}\Big)^{s}
(\vec{h}^{\transpose}\boldsymbol{\mu})^{\tilde{v}-2s}}{s!(\tilde{v}-2s)!} .
\label{eq:hermitedirect}
\end{align}
Then we can evaluate the matrix elements in 
Eq.~\eqref{eq:MEF} with the relation~\eqref{eq:hermitedirect} 
involving the summation of the 
univariate Hermite polynomials: the summation over the $s$ part in Eq.~\eqref{eq:hermitedirect} 
is the iterative expression of the univariate Hermite polynomials. 
The matrix elements of the non-Gaussian operator, for example, the position operator ($\hat{\mathbf{Q}}$) or momentum operator ($\hat{\mathbf{P}}$), can also be evaluated 
with Eqs.~\eqref{eq:hermiterec} and \eqref{eq:hermitedirect} using the following Husimi Q-functions with an auxiliary generating function parameter vector ($\boldsymbol{\lambda}$), 
\begin{align}
&\langle\boldsymbol{\alpha}\vert\hat{O}_{\mathrm{G}}\exp(\boldsymbol{\lambda}^{T}\hat{\mathbf{Q}})\vert\boldsymbol{\alpha}\rangle,\\ 
&\langle\boldsymbol{\alpha}\vert\hat{O}_{\mathrm{G}}\exp(\boldsymbol{\lambda}^{T}\hat{\mathbf{P}})\vert\boldsymbol{\alpha}\rangle,
\end{align}
because these are again complex Gaussian functions with the following identities,    
\begin{align}
    &\prod_{j=1}^{N}\hat{Q}_{j}^{n_{j}}=\prod_{j=1}^{N}\partial_{\lambda_{j}}^{n_{j}}\exp(\boldsymbol{\lambda}^{T}\hat{\mathbf{Q}})\Big\vert_{\boldsymbol{\lambda}=0}, \\ 
    &\prod_{j=1}^{N}\hat{P}_{j}^{n_{j}}=\prod_{j=1}^{N}\partial_{\lambda_{j}}^{n_{j}}\exp(\boldsymbol{\lambda}^{T}\hat{\mathbf{P}})\Big\vert_{\boldsymbol{\lambda}=0}.
\end{align}

In the subsequent section, Husimi's Q-functions are presented as complex Gaussian functions via the integral operator method~\cite{Hong-yi1994,Fan2003}.

\section{Multimode Bogoliubov integral transformation operator and Husimi's Q-function}
The most general pure Gaussian operator $\hat{O}_{\mathrm{G}}$ can be obtained as the product of the displacement and multimode squeezing operators~\cite{Ma1990,Braunstein2005}, which can be further decomposed as two rotations and a single squeezing operators. The resulting multimode Bogoliubov transformation for the bosonic annihilation ($\hat{\mathbf{a}}$, $\hat{\mathbf{b}}$) and creation ($\hat{\mathbf{a}}^{\dagger}$, $\hat{\mathbf{b}}^{\dagger}$) operators is given as follows,  
\begin{equation}
\begin{pmatrix}\hat{\mathbf{b}}\\ \hat{\mathbf{b}}^{\dagger}\end{pmatrix} = \hat{O}_{\mathrm{G}}^{\dagger}\begin{pmatrix}\hat{\mathbf{a}}\\ \hat{\mathbf{a}}^{\dagger}\end{pmatrix} \hat{O}_{\mathrm{G}} = \mathbf{K}\hat{\boldsymbol{\xi}}+\mathbf{l}  \ ,
\end{equation}
where $[\hat{a}_{j},\hat{a}_{k}^{\dagger}]=[\hat{b}_{j},\hat{b}_{k}^{\dagger}]=\delta_{jk}$, and, 
\begin{equation}
    \mathbf{K}=\begin{pmatrix}\mathbf{S} & -\mathbf{R}\\ -\mathbf{R}^{*} & \mathbf{S}^{*}\end{pmatrix},~
    \hat{\boldsymbol{\xi}}=\begin{pmatrix}\hat{\mathbf{a}}\\ \hat{\mathbf{a}}^{\dagger}\end{pmatrix},~\mathbf{l}=\begin{pmatrix} \mathbf{t}\\\mathbf{t}^{*}\end{pmatrix}. 
\end{equation}
The transformation matrix $\mathbf{K}$ satisfies the symplectic condition  
\begin{equation}
 \mathbf{K}^{\dagger}
 \begin{pmatrix}
 \mathbf{I}_{N} & \mathbf{O}_{N}\\
 \mathbf{O}_{N} & -\mathbf{I}_{N}
 \end{pmatrix}
 \mathbf{K}
 =
  \mathbf{K}
 \begin{pmatrix}
 \mathbf{I}_{N} & \mathbf{O}_{N}\\
 \mathbf{O}_{N} & -\mathbf{I}_{N}
 \end{pmatrix}
 \mathbf{K}^{\dagger}
 =
 \begin{pmatrix}
 \mathbf{I}_{N} & \mathbf{O}_{N}\\
 \mathbf{O}_{N} & -\mathbf{I}_{N}
 \end{pmatrix},
\end{equation}
and it results in the following symplectic identities, 
\begin{equation}
    \mathbf{SS}^{\dagger}-\mathbf{RR}^{\dagger}=\mathbf{I}_{N},~ \mathbf{SR}^{\mathrm{T}}=\mathbf{RS}^{\mathrm{T}}, ~
    \mathbf{S}^{\dagger}\mathbf{S}-\mathbf{R}^{\mathrm{T}}\mathbf{R}^{*}=\mathbf{I}_{N},~ \mathbf{R}^{\dagger}\mathbf{S}=\mathbf{S}^{\mathrm{T}}\mathbf{R}^{*}.
    \label{eq:sympletic}
\end{equation}
where $\mathbf{I}_{N}$ and $\mathbf{O}_{N}$ are an $N$-dimensional identity and square zero matrices, respectively.
We use, in this paper, the convention of  Ma and Rhodes~\cite{Ma1990} for  transformation of boson operator vector $\hat{\mathbf{x}}$ as
$\hat{A} \ \hat{\mathbf{x}} \ \hat{B} \equiv (\hat{A}\hat{x}_{1}\hat{B},\ldots,\hat{A}\hat{x}_{N}\hat{B})^{\mathrm{T}}$.

$\hat{O}_{\mathrm{G}}$ can be decomposed into the elementary quantum optical operators as~\cite{Braunstein2005,Huh2017VBS}, 
\begin{equation}
\hat{O}_{\mathrm{G}}=\hat{D}(\mathbf{t})\hat{R}(\mathbf{U}_{\mathrm{L}})\hat{S}(\boldsymbol{\Sigma})\hat{R}(\mathbf{U}_{\mathrm{R}}^{\dagger}) \ ,
\end{equation}
via the singular value decomposition of transformation matrices $\mathbf{S}$ and $\mathbf{R}$ (Bloch-Messiah reduction~\cite{Braunstein2005}):  $\mathbf{S}=\mathbf{U}_{\mathrm{L}}\cosh(\boldsymbol{\Sigma})\mathbf{U}_{\mathrm{R}}^{\dagger}$ and $\mathbf{R}=-\mathbf{U}_{\mathrm{L}}\sinh(\boldsymbol{\Sigma})\mathbf{U}_{\mathrm{R}}^{\mathrm{T}}$, where $\boldsymbol{\Sigma}$ is a real diagonal matrix of squeezing parameters, and $\mathbf{U}_{\mathrm{L}}$ and $\mathbf{U}_{\mathrm{R}}$ are  $N\times N$ unitary matrices. 
The elementary quantum optical unitary operators are defined as follows~\cite{Ma1990}:
\begin{align}
&\textrm{rotation operator:}\qquad\hat{R}(\mathbf{U})=\exp((\hat{\mathbf{a}}^{\dagger})^{\mathrm{T}}(\ln \mathbf{U})\hat{\mathbf{a}}),  \\
&\textrm{displacement operator:}\qquad\hat{D}(\boldsymbol{\alpha})=\exp(\boldsymbol{\alpha}^{\mathrm{T}} \hat{\mathbf{a}}^{\dagger}-\boldsymbol{\alpha}^{\dagger}\hat{\mathbf{a}}),  \\
&\textrm{squeezing operator:}\qquad\hat{S}(\Sigma)=\exp(\tfrac{1}{2}((\hat{\mathbf{a}}^{\dagger})^{\mathrm{T}}\Sigma\hat{\mathbf{a}}^{\dagger}-\hat{\mathbf{a}}^{\mathrm{T}}\Sigma^{\dagger}\hat{\mathbf{a}})). 
\end{align}
These optical operators act on $\mathbf{\hat{a}}$ as, respectively: 
\begin{align}
&\hat{R}(U)^{\dagger}\hat{\mathbf{a}}
\hat{R}(U)
= U\hat{\mathbf{a}},  \\
&\hat{D}(\boldsymbol{\alpha})^{\dagger}\mathbf{\hat{a}}\hat{D}(\boldsymbol{\alpha})=\mathbf{\hat{a}}^{\dagger}+\boldsymbol{\alpha}, 
\\
&\hat{S}(\Sigma)^{\dagger}\mathbf{\hat{a}}\hat{S}(\Sigma)=
\cosh(\Sigma)\mathbf{\hat{a}}
+\sinh(\Sigma)\mathbf{\hat{a}}^{\dagger}.
\end{align}

Fan et al.~\cite{Hong-yi1994,Fan2003} proposed an integral form of the multimode squeezing operator,   
\begin{align}
\hat{O}_{S}&=\sqrt{|\mathrm{det}(\mathbf{S})|}\int\frac{\mathrm{d}^{2}\mathbf{z}}{\pi^{N}}
\vert \mathbf{S}\mathbf{z}-\mathbf{R}\mathbf{z}^{*}\rangle\langle\mathbf{z}\vert 
\ ,
\label{eq:integraltransform0}
\end{align}
where $\mathrm{d}^{2}\mathbf{z}=\prod_{k}^{N}\mathrm{dRe}(z_{k})\mathrm{dIm}(z_{k})$, such that $\hat{\mathbf{O}}_{S}^{\dagger}\hat{\mathbf{a}}\hat{\mathbf{O}}_{S}=\mathbf{S}\hat{\mathbf{a}}-\mathbf{R}\hat{\mathbf{a}}^{\dagger}$. $\hat{\mathbf{O}}_{G}$ can also be given in an integral form by applying the displacement operator $\hat{D}(\mathbf{t})$ to the integral operator $\hat{\mathbf{O}}_{S}$, \emph{i.e.}   
\begin{align}
\hat{O}_{\mathrm{G}}&=\hat{D}(\mathbf{t})\hat{O}_{S}=\sqrt{|\mathrm{det}(\mathbf{S})|}\int\frac{\mathrm{d}^{2}\mathbf{z}}{\pi^{N}}
\hat{D}(\mathbf{t})\vert \mathbf{S}\mathbf{z}-\mathbf{R}\mathbf{z}^{*}\rangle\langle\mathbf{z}\vert, 
\nonumber \\ 
&=\sqrt{|\mathrm{det}(\mathbf{S})|}\int\frac{\mathrm{d}^{2}\mathbf{z}}{\pi^{N}}
g(\mathbf{S}, -\mathbf{R}, \mathbf{t})
\vert \mathbf{S}\mathbf{z}-\mathbf{R}\mathbf{z}^{*}+\mathbf{t}\rangle\langle\mathbf{z}\vert 
\ ,
\label{eq:integraltransform}
\end{align}
where
\begin{equation}
    g(\mathbf{S}, -\mathbf{R}, \mathbf{t})=\exp\left(\tfrac{1}{2}\boldsymbol{\xi}^{\dagger}
\begin{pmatrix}
\mathbf{S}^{\dagger}\\-\mathbf{R}^{\dagger}
\end{pmatrix}
\mathbf{t}
-\tfrac{1}{2}\boldsymbol{\xi}^{\mathrm{T}}
\begin{pmatrix}
\mathbf{S}^{\mathrm{T}}\\-\mathbf{R}^{\mathrm{T}}
\end{pmatrix}
\mathbf{t}^{*}
\right).
\end{equation}
As shown by Fan et al.~\cite{Hong-yi1994,Fan2003}, the normal ordering of $\hat{O}_{\mathrm{G}}$ is derived by using the integral form in Eq.~\eqref{eq:integraltransform} to find Husimi's Q-function. The coherent states in the integral operator are expressed as a vacuum projection form to exploit the normal ordering of the vacuum projection operator ($\vert\mathbf{0}\rangle\langle\mathbf{0}\vert$) for the integration,
\begin{align}
&g(\mathbf{S}, -\mathbf{R}, \mathbf{t})\vert \mathbf{S}\mathbf{z}-\mathbf{R}\mathbf{z}^{*}+\mathbf{t}\rangle\nonumber\\ 
&=
\exp\left(-\tfrac{1}{2}\boldsymbol{\xi}^{\mathrm{T}}\mathbf{W}\boldsymbol{\xi}+(\hat{\mathbf{a}}^{\dagger})^{\mathrm{T}}[\begin{pmatrix}\mathbf{S} &-\mathbf{R}\end{pmatrix}\boldsymbol{\xi}+\mathbf{t}]-\mathbf{t}^{\dagger}
(\mathbf{S}~-\mathbf{R})\boldsymbol{\xi}-\tfrac{1}{2}\mathbf{t}^{\dagger}\mathbf{t}+\tfrac{1}{2}\mathbf{z}^{\dagger}\mathbf{z}\right)\vert\mathbf{0}\rangle\ ,
\label{eq:ket}
\end{align}
\begin{align}
    \langle\mathbf{z}\vert=\langle\mathbf{0}\vert\exp(-\tfrac{1}{2}\mathbf{z}^{\dagger}\mathbf{z}+\hat{\mathbf{a}}^{\mathrm{T}}\mathbf{z}^{\dagger}),
\end{align}
where $\boldsymbol{\xi}^{\mathrm{T}}=(\mathbf{z}^{\mathrm{T}}~\mathbf{z}^{*\mathrm{T}})$, and with the symmetric matrix $\mathbf{W}$, 
\begin{equation}
    \mathbf{W}=
    \begin{pmatrix}-\mathbf{S}^{\mathrm{T}}\mathbf{R}^{*}&
\mathbf{S}^{\mathrm{T}}\mathbf{S}^{*}\\ \mathbf{S}^{\dagger}\mathbf{S}&-\mathbf{R}^{\mathrm{T}}\mathbf{S}^{*}
\end{pmatrix}. 
\end{equation}
By introducing the vacuum projection operator identity~\cite{louisell,Fan2003},
\begin{equation}
\vert\mathbf{0}\rangle\langle\mathbf{0}\vert=:\exp(-(\hat{\mathbf{a}}^{\dagger})^{\mathrm{T}}\hat{\mathbf{a}}): \end{equation}
where the symbol :$\hat{X}$: denotes the normal ordering of the operator $\hat{X}$, 
Eq.~\eqref{eq:integraltransform} becomes integrable,   
\begin{align}
&\hat{O}_{\mathrm{G}}=\exp(-\tfrac{1}{2}\mathbf{t}^{\dagger}\mathbf{t})\sqrt{|\mathrm{det}(\mathbf{S})|}
\nonumber \\
&\int\frac{\mathrm{d}^{2}\mathbf{z}}{\pi^{N}}
:\exp\left(-\tfrac{1}{2}\boldsymbol{\xi}^{\mathrm{T}}\mathbf{W}\boldsymbol{\xi}
-\mathbf{t}^{\dagger}
(\mathbf{S}~-\mathbf{R})\boldsymbol{\xi}
+\begin{pmatrix}(\hat{\mathbf{a}}^{\dagger})^{\mathrm{T}}\mathbf{S} &\hat{\mathbf{a}}^{\mathrm{T}}-(\hat{\mathbf{a}}^{\dagger})^{\mathrm{T}}\mathbf{R}\end{pmatrix}\boldsymbol{\xi}+(\hat{\mathbf{a}}^{\dagger})^{\mathrm{T}}\mathbf{t}-(\hat{\mathbf{a}}^{\dagger})^{\mathrm{T}}\hat{\mathbf{a}}\right):\ .
\label{eq:intordered}
\end{align}
The Gaussian integral in Eq.~\eqref{eq:intordered}, with a complex symmetric covariance matrix, can be evaluated using the following formula~\cite{Fan2003,berezin},  
\begin{align}
&\int\frac{\mathrm{d}^{2}\mathbf{z}}{\pi^{N}}
\exp\left(-\tfrac{1}{2}\boldsymbol{\xi}^{\mathrm{T}}\begin{pmatrix}\mathbf{A}&
\mathbf{B}\\ \mathbf{B}^{\mathrm{T}}&\mathbf{D}
\end{pmatrix}\boldsymbol{\xi}+\begin{pmatrix}\mathbf{u}\\ \mathbf{v}^{*}\end{pmatrix}^{\mathrm{T}}\boldsymbol{\xi}\right)
\nonumber\\
&=\left[\mathrm{det}\begin{pmatrix}\mathbf{B}^{\mathrm{T}}&\mathbf{D}\\
\mathbf{A}&\mathbf{B} 
\end{pmatrix}\right]^{-1/2}\exp\left[\frac{1}{2}\begin{pmatrix}\mathbf{u}\\ \mathbf{v}^{*}\end{pmatrix}^{\mathrm{T}}\begin{pmatrix}
\mathbf{B}^{\mathrm{T}}&\mathbf{D}\\ \mathbf{A}&\mathbf{B} 
\end{pmatrix}^{-1}\begin{pmatrix}\mathbf{v}^{*}\\ \mathbf{u}\end{pmatrix}\right],
\\ \nonumber
  &=\left[(-1)^{N}\det\begin{pmatrix}
  \mathbf{A}&\mathbf{B}\\
  \mathbf{B}^{\mathrm{T}}&\mathbf{D}
  \end{pmatrix}\right]^{-1/2}\exp\left[\frac{1}{2}\begin{pmatrix}\mathbf{u}\\ \mathbf{v}^{*}\end{pmatrix}^{\mathrm{T}}\begin{pmatrix}
  \mathbf{A}&\mathbf{B}\\\mathbf{B}^{\mathrm{T}}&\mathbf{D} 
  \end{pmatrix}^{-1}\begin{pmatrix}\mathbf{u}\\ \mathbf{v}^{*}\end{pmatrix}\right]
\ ,
\end{align}
where $\mathbf{A}$ and $\mathbf{D}$ are $N\times N$ symmetric matrices.

After integration, we find a normal ordering of the Gaussian operator, which is again a Gaussian function of the bosonic operators:  
\begin{align}
&\hat{O}_{\mathrm{G}}=\frac{\exp(-\tfrac{1}{2}\mathbf{t}^{\dagger}\mathbf{t})}{\sqrt{\vert\det(\mathbf{S})\vert}}:\exp\Big((\hat{\mathbf{a}}^{\dagger})^{\mathrm{T}}\mathbf{t}-(\hat{\mathbf{a}}^{\dagger})^{\mathrm{T}}\hat{\mathbf{a}}\nonumber \\
&+\tfrac{1}{2}\left[\hat{\boldsymbol{\xi}}^{\mathrm{T}}\begin{pmatrix}\mathbf{O}_{N} & \mathbf{I}_{N}\\ \mathbf{S}&-\mathbf{R}\end{pmatrix}-\mathbf{t}^{\dagger}
(\mathbf{S}~-\mathbf{R})\right]\mathbf{W}^{-1}\left[\begin{pmatrix}\mathbf{O}_{N} & \mathbf{S}^{\mathrm{T}}\\ \mathbf{I}_{N}&-\mathbf{R}^{\mathrm{T}}\end{pmatrix}\hat{\boldsymbol{\xi}}
-\begin{pmatrix}
\mathbf{S}^{\mathrm{T}}\\-\mathbf{R}^{\mathrm{T}}
\end{pmatrix}\mathbf{t}^{*}\right]
\Big):,
\label{eq:Oordered1}
\end{align}
where we have used the symplectic identities~\eqref{eq:sympletic} for   
\begin{equation}
(-1)^{N}\det(\mathbf{W})
=\det(\mathbf{S}^{\dagger}\mathbf{S})=\vert\det(\mathbf{S})\vert^{2}, 
\end{equation}
\begin{equation}
    \mathbf{W}^{-1}=
    \begin{pmatrix}\mathbf{R}^{\mathrm{T}}(\mathbf{S}^{\mathrm{T}})^{-1} & \mathbf{I}_{N}\\
    \mathbf{I}_{N} & (\mathbf{S}^{*})^{-1}\mathbf{R}^{*}
    \end{pmatrix}.
\end{equation}
Eq.~\eqref{eq:Oordered1} is further rearranged as
\begin{align}
&\hat{O}_{\mathrm{G}}=\frac{1}{\sqrt{\vert\det(\mathbf{S})\vert}}
:\exp\Big(\frac{1}{2}\hat{\boldsymbol{\xi}}^{\mathrm{T}}
\begin{pmatrix}
    \mathbf{R}^{\dagger}(\mathbf{S}^{\dagger})^{-1} & (\mathbf{S}^{*})^{-1}\\
    (\mathbf{S}^{\dagger})^{-1} & -(\mathbf{S}^{\dagger})^{-1}\mathbf{R}^{\mathrm{T}}
    \end{pmatrix}
    \hat{\boldsymbol{\xi}} -(\hat{\mathbf{a}}^{\dagger})^{\mathrm{T}}\hat{\mathbf{a}} \\ \nonumber 
    &+\mathbf{l}^{\mathrm{T}}
    \begin{pmatrix}
        \mathrm{0}_{N} & \mathrm{I}_{N}\\
        -(\mathbf{S}^{\dagger})^{-1}&(\mathbf{S}^{\dagger})^{-1}\mathbf{R}^{\mathrm{T}}
    \end{pmatrix}
    \hat{\boldsymbol{\xi}}
    -\frac{1}{2}\mathbf{l}^{\mathrm{T}}
    \begin{pmatrix}
        \mathbf{O}_{N}& \mathbf{O}_{N}\\
        \mathbf{I}_{N} & (\mathbf{S}^{\dagger})^{-1}\mathbf{R}^{\mathrm{T}}
    \end{pmatrix}
    \mathbf{l}
    \Big): .   
    \label{eq:Oordered2}
\end{align}

Finally, the unnormalized Husimi Q-function of the Gaussian operator is given as a complex Gaussian function,
\begin{align}
H_{\mathrm{G}}&=\exp(\boldsymbol{\alpha}^{\dagger}\boldsymbol{\alpha})\langle\boldsymbol{\alpha}\vert\hat{O}_{\mathrm{G}}\vert\boldsymbol{\alpha}\rangle\\ \nonumber
&=\langle\mathbf{0}\vert\hat{O}_{G}\vert\mathbf{0}\rangle
\exp\Big(-\frac{1}{2}\bar{\boldsymbol{\alpha}}^{\mathrm{T}}
\mathbf{V}    
\bar{\boldsymbol{\alpha}}+
    \boldsymbol{\mu}^{\mathrm{T}}
    \bar{\boldsymbol{\alpha}}
    \Big),
    \label{eq:wHG}
\end{align}
where $\bar{\boldsymbol{\alpha}}^{\mathrm{T}}=(\boldsymbol{\alpha}^{\mathrm{T}} \boldsymbol{\alpha}^{\dagger})$ and 
\begin{align}
    &\mathbf{V}=
\begin{pmatrix}
    -\mathbf{R}^{\dagger}(\mathbf{S}^{\dagger})^{-1} & -(\mathbf{S}^{*})^{-1}\\
    -(\mathbf{S}^{\dagger})^{-1} & (\mathbf{S}^{\dagger})^{-1}\mathbf{R}^{\mathrm{T}}
    \end{pmatrix},\\ 
    &\boldsymbol{\mu}^{\mathrm{T}}=\mathbf{l}^{\mathrm{T}}
    \begin{pmatrix}
        \mathrm{0}_{N} & \mathrm{I}_{N}\\
        -(\mathbf{S}^{\dagger})^{-1}&(\mathbf{S}^{\dagger})^{-1}\mathbf{R}^{\mathrm{T}}
    \end{pmatrix},\\ 
        &\langle\mathbf{0}\vert\hat{O}_{G}\vert\mathbf{0}\rangle=\frac{1}{\sqrt{\vert\det(\mathbf{S})\vert}}\exp\left[-\frac{1}{2}\mathbf{l}^{\mathrm{T}}
    \begin{pmatrix}
        \mathbf{O}_{N}& \mathbf{O}_{N}\\
        \mathbf{I}_{N} & (\mathbf{S}^{\dagger})^{-1}\mathbf{R}^{\mathrm{T}}
    \end{pmatrix}
    \mathbf{l}\right].
\end{align}
Eq.~\eqref{eq:wHG} agrees with the expression of Dondonov and Man'ko~\cite{dondonov1994}, which was obtained by a complex linear canonical transformation~\cite{malkin1973,wolf1974}.

Alternatively, we can obtain the unnormalrized Husimi Q-function via the integral transformation directly without invoking the normally ordered form, that is,  
\begin{align}
&\exp(\boldsymbol{\alpha}^{\dagger}\boldsymbol{\alpha})\langle\boldsymbol{\alpha}\vert\hat{O}_{\mathrm{G}}\vert\boldsymbol{\alpha}\rangle=\exp(\boldsymbol{\alpha}^{\dagger}\boldsymbol{\alpha})\sqrt{|\mathrm{det}(\mathbf{S})|}\int\frac{\mathrm{d}^{2}\mathbf{z}}{\pi^{N}}g(\mathbf{S}, -\mathbf{R}, \mathbf{t})\langle\boldsymbol{\alpha}\vert \mathbf{S}\mathbf{z}-\mathbf{R}\mathbf{z}^{*}+\mathbf{t}\rangle\langle\mathbf{z}\vert\boldsymbol{\alpha}\rangle,  \nonumber \\ 
&=\exp(-\tfrac{1}{2}\mathbf{t}^{\dagger}\mathbf{t})
\exp(\boldsymbol{\alpha}^{\dagger}\mathbf{t})
\sqrt{|\mathrm{det}(\mathbf{S})|} 
\nonumber\\ 
&~~~\int\frac{\mathrm{d}^{2}\mathbf{z}}{\pi^{N}}
\exp\left(-\tfrac{1}{2}\boldsymbol{\xi}^{\mathrm{T}}\mathbf{W}\boldsymbol{\xi}+
\left[(\bar{\boldsymbol{\alpha}})^{\mathrm{T}}
\begin{pmatrix}
\mathbf{O}_{N}&\mathbf{I}_{N}\\
\mathbf{S}&-\mathbf{R}
\end{pmatrix}
-\mathbf{t}^{\dagger}
(\mathbf{S}~-\mathbf{R})
\right]\boldsymbol{\xi}
\right), \nonumber \\ 
&=\frac{\exp(-\tfrac{1}{2}\mathbf{t}^{\dagger}\mathbf{t})}{\sqrt{\vert\det(\mathbf{S})\vert}}
\exp(\boldsymbol{\alpha}^{\dagger}\mathbf{t})\nonumber \\ 
&~~~\exp\left(\frac{1}{2}\left[\bar{\boldsymbol{\alpha}}^{\mathrm{T}}\begin{pmatrix}\mathbf{O}_{N} & \mathbf{I}_{N}\\ \mathbf{S}&-\mathbf{R}\end{pmatrix}-\mathbf{t}^{\dagger}
(\mathbf{S}~-\mathbf{R})\right]\mathbf{W}^{-1}\left[\begin{pmatrix}\mathbf{O}_{N} & \mathbf{S}^{\mathrm{T}}\\ \mathbf{I}_{N}&-\mathbf{R}^{\mathrm{T}}\end{pmatrix}\bar{\boldsymbol{\alpha}}-
\begin{pmatrix}
\mathbf{S}^{\mathrm{T}}\\-\mathbf{R}^{\mathrm{T}}
\end{pmatrix}\mathbf{t}^{*}\right]\right),\nonumber \\ 
&=H_{\mathrm{G}}
\end{align}

We can exploit the direct integral transformation method to derive non-Gaussian matrix elements easily in the MHP formula. Here, we derive the matrix elements of the powers of position 
($\hat{\mathbf{Q}}=\tfrac{1}{\sqrt{2}}(\hat{\mathbf{a}}+\hat{\mathbf{a}}^{\dagger})$) and momentum ($\hat{\mathbf{P}}=\tfrac{-\mathrm{i}}{\sqrt{2}}(\hat{\mathbf{a}}-\hat{\mathbf{a}}^{\dagger})$) operators. 
Using the following identities~\cite{Fan2003}, we express the exponential operators of position and momentum in the ordered form, 
\begin{align}
    &\int_{-\infty}^{\infty}\mathrm{d}\mathbf{q}\vert\mathbf{q}\rangle\langle\mathbf{q}\vert=\int_{-\infty}^{\infty}\frac{\mathrm{d}\mathbf{q}}{\sqrt{\pi}^{N}}:\exp(-(\mathbf{q}-\hat{\mathbf{Q}})^{2}):=1, \\ 
    &\int_{-\infty}^{\infty}\mathrm{d}\mathbf{p}\vert\mathbf{p}\rangle\langle\mathbf{p}\vert=\int_{-\infty}^{\infty}\frac{\mathrm{d}\mathbf{p}}{\sqrt{\pi}^{N}}:\exp(-(\mathbf{p}-\hat{\mathbf{P}})^{2}):=1,    
\end{align}
where $\mathbf{q}$ and $\mathbf{p}$ are the position and momentum vectors, respectively, as 
\begin{align}
    &\exp(\boldsymbol{\lambda}^{T}\hat{\mathbf{Q}})=:\exp(\boldsymbol{\lambda}^{T}\hat{\mathbf{Q}}+\frac{\boldsymbol{\lambda}^{\mathrm{T}}\boldsymbol{\lambda}}{4}):=\exp(\frac{\boldsymbol{\lambda}^{\mathrm{T}}\boldsymbol{\lambda}}{4}):\exp(\frac{\boldsymbol{\lambda}^{\mathrm{T}}}{\sqrt{2}}(\mathbf{I}_{N}~\mathbf{I}_{N})\hat{\boldsymbol{\xi}}):,\label{eq:coonormal}\\
    &\exp(\boldsymbol{\lambda}^{T}\hat{\mathbf{P}})=:\exp(\boldsymbol{\lambda}^{T}\hat{\mathbf{P}}+\frac{\boldsymbol{\lambda}^{\mathrm{T}}\boldsymbol{\lambda}}{4}):=\exp(\frac{\boldsymbol{\lambda}^{\mathrm{T}}\boldsymbol{\lambda}}{4}):\exp(\frac{-\mathrm{i}\boldsymbol{\lambda}^{\mathrm{T}}}{\sqrt{2}}(\mathbf{I}_{N}~-\mathbf{I}_{N})\hat{\boldsymbol{\xi}}):. \label{eq:momnormal}
\end{align}
It is now straightforward to find the unnormalized Husimi Q-function for the coordinate operator with the normally ordered form of the exponential coordinate operator~\eqref{eq:coonormal} (a similar calculation can be applied to the exponential momentum operator~\eqref{eq:momnormal}) and the integral transformation technique; the result is 
\begin{align}
&\exp(\boldsymbol{\alpha}^{\dagger}\boldsymbol{\alpha})\langle\boldsymbol{\alpha}\vert\hat{O}_{\mathrm{G}}\exp(\boldsymbol{\lambda}^{T}\hat{\mathbf{Q}})\vert\boldsymbol{\alpha}\rangle, \nonumber\\
&=\exp(\boldsymbol{\alpha}^{\dagger}\boldsymbol{\alpha})\sqrt{|\mathrm{det}(\mathbf{S})|}\int\frac{\mathrm{d}^{2}\mathbf{z}}{\pi^{N}}g(\mathbf{S}, -\mathbf{R}, \mathbf{t})\langle\boldsymbol{\alpha}\vert \mathbf{S}\mathbf{z}-\mathbf{R}\mathbf{z}^{*}+\mathbf{t}\rangle\langle\mathbf{z}\vert\exp(\boldsymbol{\lambda}^{T}\hat{\mathbf{Q}})\vert\boldsymbol{\alpha}\rangle, \nonumber\\
&=\exp(-\tfrac{1}{2}\mathbf{t}^{\dagger}\mathbf{t})\exp(\tfrac{1}{4}\boldsymbol{\lambda}^{\mathrm{T}}\boldsymbol{\lambda})
\exp\left((\bar{\boldsymbol{\alpha}})^{\mathrm{T}}
\begin{pmatrix}
\tfrac{1}{\sqrt{2}}\boldsymbol{\lambda}\\
\mathbf{t}
\end{pmatrix}
\right)
\sqrt{|\mathrm{det}(\mathbf{S})|}
\nonumber \\ 
&~~~\int\frac{\mathrm{d}^{2}\mathbf{z}}{\pi^{N}}
\exp\left(-\tfrac{1}{2}\boldsymbol{\xi}^{\mathrm{T}}\mathbf{W}\boldsymbol{\xi}+
\left[(\bar{\boldsymbol{\alpha}})^{\mathrm{T}}
\begin{pmatrix}
\mathbf{O}_{N}&\mathbf{I}_{N}\\
\mathbf{S}&-\mathbf{R}
\end{pmatrix}
-\mathbf{t}^{\dagger}
(\mathbf{S}~-\mathbf{R})+(\mathbf{0}^{\mathrm{T}}~\tfrac{1}{\sqrt{2}}\boldsymbol{\lambda}^{\mathrm{T}})
\right]\boldsymbol{\xi}
\right),\nonumber \\ 
&=\footnotesize{\frac{\exp(-\tfrac{1}{2}\mathbf{t}^{\dagger}\mathbf{t})}{\sqrt{\vert\det(\mathbf{S})\vert}}
\exp(\tfrac{1}{4}\boldsymbol{\lambda}^{\mathrm{T}}\boldsymbol{\lambda})
\exp\left((\bar{\boldsymbol{\alpha}})^{\mathrm{T}}
\begin{pmatrix}
\tfrac{1}{\sqrt{2}}\boldsymbol{\lambda}\\
\mathbf{t}
\end{pmatrix}
\right)}
\nonumber\\
&\footnotesize{\exp\left(\tfrac{1}{2}\left[\bar{\boldsymbol{\alpha}}^{\mathrm{T}}\begin{pmatrix}\mathbf{O}_{N} & \mathbf{I}_{N}\\ \mathbf{S}&-\mathbf{R}\end{pmatrix}-\mathbf{t}^{\dagger}
(\mathbf{S}~-\mathbf{R})+(\mathbf{0}^{\mathrm{T}}~\tfrac{1}{\sqrt{2}}\boldsymbol{\lambda}^{\mathrm{T}})\right]\mathbf{W}^{-1}\left[\begin{pmatrix}\mathbf{O}_{N} & \mathbf{S}^{\mathrm{T}}\\ \mathbf{I}_{N}&-\mathbf{R}^{\mathrm{T}}\end{pmatrix}\bar{\boldsymbol{\alpha}}-
\begin{pmatrix}\mathbf{S}^{\mathrm{T}}\\-\mathbf{R}^{\mathrm{T}}\end{pmatrix}\mathbf{t}^{*}
+\begin{pmatrix}
\mathbf{0}\\\tfrac{1}{\sqrt{2}}\boldsymbol{\lambda}
\end{pmatrix}
\right]\right)}.
\end{align}
The above Gaussian function is further rearranged as follows: 
\begin{align}
&\exp(\boldsymbol{\alpha}^{\dagger}\boldsymbol{\alpha})\langle\boldsymbol{\alpha}\vert\hat{O}_{\mathrm{G}}\exp(\boldsymbol{\lambda}^{T}\hat{\mathbf{Q}})\vert\boldsymbol{\alpha}\rangle\nonumber\\ 
&=\langle\mathbf{0}\vert\hat{O}_{G}\vert\mathbf{0}\rangle
\exp\left(-\tfrac{1}{2}
\begin{pmatrix}
\bar{\boldsymbol{\alpha}} \\ \boldsymbol{\lambda}
\end{pmatrix}^{\mathrm{T}}
\bar{\mathbf{V}}
\begin{pmatrix}
\bar{\boldsymbol{\alpha}} \\ \boldsymbol{\lambda}
\end{pmatrix}
+\bar{\boldsymbol{\mu}}^{\mathrm{T}}
\begin{pmatrix}
\bar{\boldsymbol{\alpha}} \\ \boldsymbol{\lambda}
\end{pmatrix}
\right), 
\end{align}
where 
\begin{equation}
\small{    \bar{\mathbf{V}}=
    \begin{pmatrix}
\mathbf{V} & 
-\tfrac{1}{\sqrt{2}}\begin{bmatrix}(\mathbf{S}^{*})^{-1}\mathbf{R}^{*}+\mathbf{I}\\(\mathbf{S}^{\dagger})^{-1}\end{bmatrix}\\
-\tfrac{1}{\sqrt{2}}\begin{bmatrix}(\mathbf{S}^{*})^{-1}\mathbf{R}^{*}+\mathbf{I}\\(\mathbf{S}^{\dagger})^{-1}\end{bmatrix}^{\mathrm{T}}& 
-\tfrac{1}{2}((\mathbf{S}^{*})^{-1}\mathbf{R}^{*}+\mathbf{I})
\end{pmatrix},~ 
\bar{\boldsymbol{\mu}}^{\mathrm{T}}=
\begin{pmatrix}
\boldsymbol{\mu}^{\mathrm{T}}& -\tfrac{1}{2\sqrt{2}}\mathbf{t}^{\dagger}(\mathbf{S}^{\dagger})^{-1}
\end{pmatrix}
}.
\end{equation}
Therefore, the matrix elements of Gaussian/non-Gaussian operators in the Fock state basis can be obtained using the MHP formula~\eqref{eq:MEF}. 

\section{Conclusions}
In this paper, we presented the connection between the multivariate Hermite polynomials and the multivariate Gaussian moments, where the matrix elements of Gaussian/non-Gaussian operators in the Fock state basis can be evaluated with the existing algorithms of the MHPs and MGMs. By adapting the integral transformation operator method of Fan et al.~\cite{Hong-yi1994,Fan2003} for the multimode Bogoliubov transformation, the generating functions (Husimi's Q-function) for the Gaussian/non-Gaussian operator matrix elements are easily derived. The method introduced in this paper may be useful for developing numerical schemes or quantum optical circuits involving Gaussian states.            

\section*{Acknowledgements}
This work was supported by Basic Science Research Program through the National Research Foundation of Korea (NRF) funded by the Ministry of Education, Science and Technology (NRF-2015R1A6A3A04059773, NRF-2019M3E4A1080227, NRF-2019M3E4A1079666). The author also acknowledges the support by POSCO Science Fellowship of POSCO TJ Park Foundation.  

\section*{References}
\providecommand{\newblock}{}

\end{document}